\documentstyle[twocolumn,aps]{revtex}
\begin{document}
\draft
\title{\bf Is the Peak Value of $\sigma_{xx}$ at the Quantum Hall Transition
Universal?}
\author{K. Ziegler$^{a,b}$ and G. Jug$^a$}
\address{$^a$Max-Planck-Institut f\"ur Physik Komplexer Systeme,
Au\ss enstelle Stuttgart, Postfach 800665, D-70506 Stuttgart,
Germany\\
$^b$Institut f\"ur Physik, Universit\"at Augsburg, D-86135 Augsburg, Germany}
\date{\today}
\maketitle 
\begin{abstract}
The question of the universality of the longitudinal peak conductivity at 
the integer quantum Hall transition is considered. For this purpose,
a system of 2D Dirac fermions with random mass characterised by variance $g$ 
is proposed as a model which undergoes a quantum Hall transition. Whilst
for some specific models the longitudinal peak conductivity $\sigma_{xx}$
was found to be universal (in agreement with the conjecture of Lee et al.
as well as with some numerical work), we find that $\sigma_{xx}$ is reduced
by a factor $(1+g/2\pi)^{-1}$, at least for small $g$. This provides some
theoretical evidence for the non-universality of $\sigma_{xx}$, as observed
in a number of experiments.
\end{abstract}


The integer quantum Hall effect is characterized by the very accurate
and robust plateaux of the Hall conductivity $\sigma_{xy}=ne^2/h$, where
$n$ is an integer \cite{nobel}. It is a common belief that the accuracy of the 
plateaux is a consequence of the fact that the Hall current at the plateaux
is carried by edge states \cite{halperin}. This picture is also supported by 
the observation that the longitudinal conductivity $\sigma_{xx}$ vanishes at 
the Hall plateaux, since this is a dissipative quantity which requires 
extended bulk states.

At the transition between two Hall plateaux, on the other hand, $\sigma_{xx}$
becomes nonzero, indicating that extended bulk states are created. It has been
conjectured in the literature that the maximum of $\sigma_{xx}$ has
also a universal value in units of $e^2/h$ like $\sigma_{xy}$, independent
of the specific properties of the sample \cite{lee0,huo,lee}. Specifically,
a universal value was found in a number of numerical studies of the lowest
Landau level projected electrons \cite{huo,gammel} and of the 
Chalker--Coddington network model \cite{lee}. Similarly, a universal value 
of $e^2/\pi h$ was found through investigations based on field theory. For
instance, Hikami et al. \cite{hikami} obtained this value in a model for 
electrons at the lowest Landau level with random spin-scattering. The same 
value was also found later for Dirac fermions with a random vector potential 
\cite{ludwigetal}, independently of the strength of randomness, and for Dirac 
fermions with a weakly-random mass \cite{zie6}. This is remarkable, because 
Dirac fermions describe the Hall plateaux as well as the transition
between Hall plateaux \cite{ludwigetal,zie6,zie2}.

Experimentally, the value of $\sigma_{xx}^{max}$ is much less robust than the 
Hall plateaux and varies between 0.2 and 0.5 in units of $e^2/h$ 
\cite{wei,yamane,mceuen,shahar,rokhinson}. In particular, in a recent
experiment it was found that $\sigma_{xx}^{max}$ presents for two different
samples, at the same low temperature and filling factor, variations as 
important as 40 \%  \cite{rokhinson}.

From a fundamental point of view it appears to be rather unlikely that the 
dissipative conductivity $\sigma_{xx}$ is so robust so as to be unaffected 
by the properties of the material, e.g. by impurities. This problem was 
discussed recently by Ruzin et al. \cite{ruzin} and Cooper  et al. 
\cite{cooper} using percolation theory for the random carrier distribution 
in a disordered sample. They concluded from their calculations that there 
cannot be a universal conductivity peak height. This raises the question 
as to whether the models studied in \cite{hikami,ludwigetal} describe a 
generic situation (such as in experiments) or only a very special case in 
which the effect of disorder is restricted due to special conservation laws.
Indeed, it was already discussed in Ref. \cite{ludwigetal} that a random vector
potential is not a generic case since randomness in the vector potential does 
not create electronic states at the Hall transition. The existence of these
states, however, is necessary in order to describe a realistic situation. 
Therefore, a generic model is probably more like a combination of a random 
vector potential as well as a random Dirac mass. This view is also supported 
by the Chalker--Coddington network model which is in the large--scale limit 
equivalent to Dirac fermions with random vector potential, random Dirac mass 
and random energy \cite{ho}.

The simplest model with non-vanishing density of states at the Hall transition
is given by 2D Dirac fermions with a random mass. Although this does not 
represent the most general case, it may be appropriate to investigate the 
effect of more realistic randomness on the maximum value of $\sigma_{xx}$ at 
the Hall transition. For this purpose we will extend the evaluation of 
$\sigma_{xx}^{max}$, performed by one of the authors for a weakly random 
Dirac mass \cite{zie6}, to stronger randomness. 

The 2D Dirac fermion model is defined by the Hamiltonian 
\cite{ludwigetal,zie6,zie2}

\begin{equation}
H_D=M\sigma_3+i\nabla_1\sigma_1+i\nabla_2\sigma_2,
\end{equation}

\noindent
where $\sigma_j$ are Pauli matrices, $\nabla_j$ the gradient in $j$-direction
and $M$ a random mass with mean $m$ and correlation 
$\langle M_rM_{r'}\rangle=g\delta_{rr'}$.
The longitudinal conductivity $\sigma_{xx}$ can be evaluated from the
two--particle Green's function 
$C(r,\eta)=\sum_{jj'}\langle |(H_D+i\eta)^{-1}_{jj',r0}|^2\rangle$
\cite{ludwigetal} which connects the
origin $r=0$ of the two--dimensional electron gas with the site $r$. 
$\langle ...\rangle$ refers to the average over the random Dirac mass.
According to Ref. \cite{zie6} the Fourier components of $C(r,\eta)$ are given
by a function of the 2D wave vector $k$ and the frequency $\eta$

\begin{equation}
{\tilde C}(k,\eta)
=(\eta'/2g)[\eta+(g\eta'D'/2)k^2]^{-1}.
\label{cricor'}
\end{equation}

\noindent
That is, the average two-particle Green's function describes a diffusion 
process with diffusion coefficient $D=g\eta'D'/2$, where $D'$ is given by

\begin{equation}
D'=4\alpha\Big[
1+\alpha({ \mu^2\over1/g-2\alpha \mu^2}+{ {\mu^*}^2\over1/g-2\alpha
{\mu^*}^2})\Big]
\label{diff0'}
\end{equation}

\noindent
with $\mu=m'+i\eta'$ and

\begin{equation}
\alpha=\int(|\mu|^2+k^2)^{-2}d^2k/4\pi^2\sim 
{1\over4\pi({m'}^2+{\eta'}^2)}.
\end{equation}

\noindent
The parameters $\mu'$ and $\eta'$ were evaluated in 
saddle point approximation \cite{zie6}. They obey the following equations

\begin{equation}
\eta'-\eta =\eta' g I\ \ \ {\rm and}\ m'=m/(1+gI)
\label{spe}
\end{equation}

\noindent
with $I\sim -{1\over \pi}\ln|\mu|$. 

The longitudinal conductivity is directly connected with the diffusion
coefficient via the Einstein relation

\begin{equation}
\sigma_{xx}={e^2\over\hbar}D\rho,
\end{equation}

\noindent
where $\rho$ is the density of states. The latter is given in the model under
consideration as $\rho=\eta'/\pi g$,
and $D'$ can be approximated in (\ref{diff0'}) as $D'\approx 4\alpha$ for
very weak disorder ($g\approx0$) \cite{zie6}.
This yields a peak conductivity $\sigma_{xx}^{max}
\approx e^2/\pi h$. In the following we will extend this result to the case of
stronger
disorder by taking the full expression of $D'$ in (\ref{diff0'}). Although
this requires a complicated calculation for the $m$--dependent conductivity
in general,
we will find a surprisingly simple result if only the peak value is considered.

In principle we can evaluate the parameters $\mu'$ and $\eta'$ by solving
Eq. (\ref{spe}) in order to obtain $\sigma_{xx}$.
However, this is not necessary if we use the fact that the maximum of
$\sigma_{xx}$ is at $m=0$ \cite{zie6}, the point where the average
mass of the Dirac fermions vanishes.
In this case we have $\alpha=1/4\pi{\eta'}^2$ and

\begin{equation}
D'={1\over\pi{\eta'}^2}{1\over1+g/2\pi}.
\label{D'0'}
\end{equation}

\noindent
Moreover, for the diffusion coefficient this yields immediately 
$D=(g\eta'/2)D'=1/2\pi^2\rho(1+g/2\pi)$ and eventually for
the peak conductivity

\begin{equation}
\sigma^{max}_{xx}={e^2\over h\pi}{1\over 1+g/2\pi}.
\end{equation}

\noindent
Thus the maximum of $\sigma^{max}_{xx}$ depends on the strength of the random
fluctuations $g$. The universal value of Dirac fermions with random vector
potential agrees only when the variance of the Dirac mass vanishes, $g=0$,
that is for the pure system. For real samples, however, $g$ might be small 
(e.g., $0.1$ in units of the hopping energy of the electrons) such that the 
reduction of the peak height is small. This is in good agreement with recent 
experiments which found $\sigma_{xx}^{max}\approx 0.2 \cdots 0.35 e^2/h$ 
for different samples \cite{rokhinson}. For very strong
randomness $g\gg 1$ there may be a transition to a Hall insulator 
\cite{shahar} and our Dirac fermion model would not be probably sufficient to
describe this behaviour adequately.

In conclusion, we have shown that generic randomness in the 2D Dirac fermion
model for the integer quantum Hall transition displays no universality in
the peak value of the longitudinal conductivity. Our calculation supports the
findings by Ruzin et al. \cite{ruzin} and Cooper et al. \cite{cooper}, as
well as the various experimental observations in the literature. Moreover,
our value of $\sigma_{xx}^{max}$ is in good agreement with the experimental
result of Rokhinson et al. \cite{rokhinson}.

\end{document}